# Energy Transport Between Hole Gas and Crystal Lattice in Diluted Magnetic Semiconductor

J.M. Kivioja[1], M. Prunnila[1], S. Novikov[2], P. Kuivalainen[2], and J. Ahopelto[1]

[1]*VTT Micro and Nanoelectronics, Tietotie 3, Espoo, P.O. Box 1000, FI-02044 VTT, Finland*
[2]*Electron Physics Laboratory, Helsinki University of Technology, P.O.BOX 3500, FI-02015 HUT, Finland*

**Abstract.** The temperature dependent energy transfer rate between hole gas and lattice has been investigated in thin $Mn_xGa_{1-x}As$ (x=3.7% and 4.0%) films by heating the hole system with power density $P_d$ and measuring the hole temperature $T$. The heating experiments were carried out in temperature range of 250 mK-1.3 K and the temperature dependency of resistivity provided the hole thermometer. When the hole temperature greatly exceeds the lattice temperature we find that $P_d \sim T^n$, where $n$ is in the range of 4 - 5.

## INTRODUCTION

Recent years have seen a growing interest in the field of diluted magnetic semiconductors due to their ability to combine magnetism and integrated electronics. Various electrical and material properties of magnetic semiconductors have been extensively reported in the literature. Here, we focus to study thermal relaxation of holes in magnetic semiconductor, which is a new subject in this field. In this work the temperature dependent energy loss rate between charge carriers and lattice is experimentally investigated in MnGaAs.

## SAMPLES

The MnGaAs samples were grown on semi-insulating GaAs (100) substrate by molecular beam epitaxy (MBE). First an undoped GaAs buffer layer (230 nm) was grown at 580 °C. Then low temperature MBE of a 100 nm thick $Mn_xGa_{1-x}As$ device layer was performed at 230 °C. In the experiments we studied two samples with manganese concentrations of x=3.7 % and 4.0 % and Curie temperatures of 60 K and 62 K, respectively.

The MnGaAs films were patterned utilizing UV-lithography and wet etching. The MnGaAs mesas were contacted with superconducting electrodes (Nb or Al) for preventing thermal leakage during the heating experiments. Samples were electrically characterized down to 250 mK using $He^3$-sorption cryostat and standard lock-in techniques.

## RESULTS AND DISCUSSION

Figure 1 presents measured resistivity $\rho$ as a function of temperature. The resistivity of both samples show a peak around 60 K. The position of the peak gives the Curie temperature. Furthermore, the Hall-resistivities $\rho_{xy}$ (not shown) exhibit the standard anomalous Hall effect: at low magnetic fields $B$ (<0.4 T) $\rho_{xy}$ shows a rabid increase with increasing $B$ and then it almost saturates at $B \sim 0.4$ T. Additionally, $\rho_{xy}(B)$ shows hysteretic behavior. The observed features in $\rho$ and $\rho_{xy}$ are well known signatures of ferromagnetism [1].

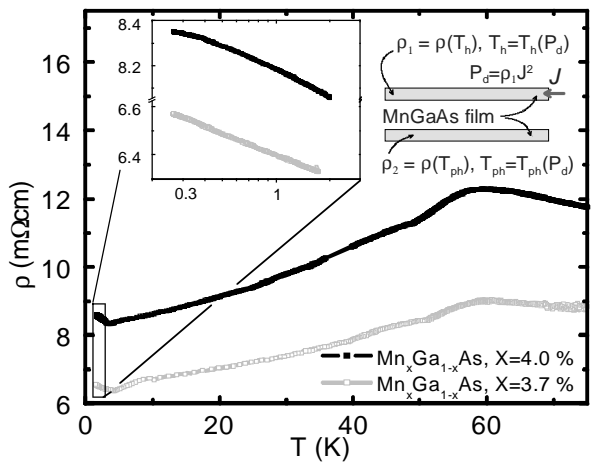

**FIGURE 1.** Temperature dependency of the resistivity $\rho$. The right inset shows the schematic top view of the heating experiments and also how $\rho(T)$ is utilized in determining the hole and phonon temperatures (see Fig. 2 for the results).

Below T ~ 5 K both samples exhibit a "Kondo-like" $\rho(T) \sim \log(T)$ behavior (left inset in Fig. 1) [2]. This well defined temperature dependency of the resistivity provides a local thermometer, which is utilized in the heating experiments: the resistivity of adjacent electronically isolated films gives the hole $T_h$ and phonon temperatures $T_{ph}$ in the spirit of Refs. [3,4] (see the right inset of Fig. 1). The thermometers were first calibrated by slowly adjusting the bath temperature of the cryostat. Then one of the 5 μm wide adjacent MnGaAs mesas was heated by applying a DC current density $J$ at constant bath (cryostat) temperature. The change in the resistivities, which were measured with a small ac-signal, give the response of $T_h$ and $T_{ph}$ to heating power density $P_d = \rho J^2$. These responses are plotted in Fig. 2 at various bath temperatures.

The temperatures of electrically isolated phonon and hole thermometers have very different responses. The hole temperature shows a strong response while the phonon temperature shows only extremely weak increase, which is observable at the highest heating powers and lowest bath temperatures. This indicates that the thermal coupling between GaAs substrates and cryostat's sample holder (copper) was extremely good, contradictory to studies utilizing silicon substrates [3,4]. Further, this also shows that the "bottle-neck" in the heat path is the hole-phonon energy relaxation rate.

Figure 2 shows also curves $P_d \propto T^n$ with n = 4,5. We can observe that the experimental $T_h$ fall between these dependencies at high power density. More careful inspection reveals that the sample with x = 3.7 % (4.0 %) is described better with n=4 (n=5). On the other hand, the sensitivity of our hole (and phonon) thermometer is rather limited, which makes clear distinction between these two power laws difficult. Thus, we conclude that our preliminary experiments indicate that $P_d \propto T^n$ with n = 4 - 5 for both x = 3.7 % and x = 4.0 %. Note that as the hole-phonon energy relaxation rate is given by $\tau_{h-ph}^{-1} \propto T^{n-2}$ [4] our results also suggest that in $Mn_xGa_{1-x}As$ $\tau_{h-ph}^{-1} \propto T^{2-3}$.

In semiconductors the elastic intra-valley [5] and inter-valley [4] scattering processes affect the carrier-phonon relaxation at low temperatures. In low mobility hole systems the latter is expect to have a strong effect, because it does not require diffusion. The role of ferromagnetism in the energy relaxation is not yet clear. However, we have also performed heating measurements at various perpendicular magnetic fields between -2 T and 2 T and our conclusion is that the hole-phonon relaxation rate in $Mn_xGa_{1-x}As$ has negligible magnetic field dependency.

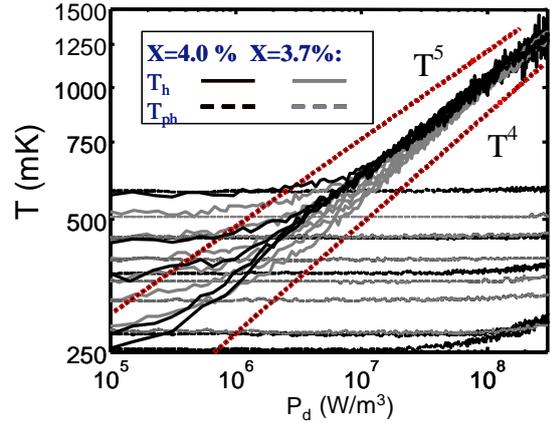

**FIGURE 2.** Hole-gas temperature ($T_h$) and temperature of the lattice ($T_{ph}$) as a function of heating power density at various base temperatures.

## SUMMARY

Our results show that the energy loss rate from charge carriers to lattice in $Mn_xGa_{1-x}As$ samples with x=3.7 % and x=4.0 % follows a power law $P_d \propto T^{4-5}$, which indicates that the hole-phonon energy relaxation time has $\tau_{h-ph}^{-1} \propto T^{2-3}$ behavior. We observed no magnetic field dependency in the energy relaxation.

## ACKNOWLEDGMENTS

The Academy of Finland (grant 205478) is acknowledged for financial support.